\documentclass[11pt]{article}
\usepackage[dvips]{graphicx} 
\usepackage{latexsym}
\usepackage[authoryear]{natbib}
\let \chapter \section
\usepackage{enumerate}
\usepackage{multirow}
\usepackage{xfrac}
\usepackage{lmodern}
\usepackage{epsfig}
\usepackage{graphics,graphicx,color}
\usepackage{amssymb,amsfonts,amsthm,amsmath}
\usepackage{multirow, booktabs}
\usepackage{mathrsfs}
\usepackage{subfig} 

\newcommand{\diag}{\operatorname{diag}}
\newcommand{\vectorize}{\operatorname{vec}}

\usepackage{amsmath,bbm,amssymb,mathrsfs}
\usepackage{amsthm,amscd,float}             
\usepackage{graphicx, multirow, graphics}
\usepackage{comment}
\usepackage{longtable, lscape, setspace}  
\usepackage{rotating}
\usepackage{dcolumn}
\usepackage{natbib}
\usepackage{booktabs,caption,fixltx2e}
\usepackage[flushleft]{threeparttable}
\usepackage{caption}
\usepackage{hyperref}

\newtheorem{algorithm}{Algorithm}

\newcommand{\bm}[1]{\mbox{\boldmath$ #1 $\unboldmath}}

\def \diag{{\rm diag}}

\title{Multivariate Regression of Mixed Responses for Evaluation of Visualization Designs} 

\author{Xiaoning Kang$^{1}$, Xiaoyu Chen$^{2}$, Ran Jin$^{2}$, Hao Wu$^{3}$ and Xinwei Deng$^{4}$ \\
$^{1}$International Business College,
Dongbei University of Finance and Economics, China \\
$^{2}$ Grado Department of Industrial and Systems Engineering,
Virginia Tech, USA \\
$^{3}$Discovery Analytics Center, Virginia Tech, USA \\
$^{4}$Department of Statistics, Virginia Tech, USA  \\
Correspondence: Xinwei Deng, Associate Professor, email: xdeng@vt.edu.} 

\begin{document}

\maketitle 

\begin{abstract}
Information visualization significantly enhances human perception by graphically representing complex data sets.
The variety of visualization designs makes it challenging to efficiently evaluate all possible designs catering to users' preferences and characteristics.
Most of existing evaluation methods perform user studies to obtain multivariate qualitative responses from users via questionnaires and interviews.
However, these methods cannot support online evaluation of designs as they are often time-consuming.
A statistical model is desired to predict users' preferences on visualization designs based on non-interference measurements (i.e., wearable sensor signals).
In this work, we propose a multivariate regression of mixed responses (MRMR) to facilitate quantitative evaluation of visualization designs.
The proposed MRMR method is able to provide accurate model prediction with meaningful variable selection.
A simulation study and a user study of evaluating visualization designs with 14 effective participants are conducted to illustrate the merits of the proposed model.
\end{abstract}


\section{Introduction}
\label{sec:intro}
Comprehending complex data sets is a cognitive challenge yet important task for human-machine collaborations.
To reduce the mental workload and to facilitate fast comprehension of data, information visualization techniques have been widely applied to graphically represent complex data sets using different visualization designs.
For example, visualization techniques have been adopted for visualizing manufacturing simulations (\citealt{rohrer2000seeing}), sequence management (\citealt{sackett2006review}), and virtual design of factories (\citealt{lindskog2013visualization}).
As a mix-up of reality and virtual objects, \cite{chen2016iserc} proposed an augmented reality (AR)-based visualization platform to visualize real-time data streams online and incorporated an online quality-process model for a fused deposition modeling process.
Besides, AR-based and virtual reality (VR)-based visualization systems have also been developed for various applications in human-machine collaboration, such as facility planning (\citealt{dangelmaier2005virtual}), production control (\citealt{damiani2018augmented}), and workforce training (\citealt{wang2017augmented}).
With a large amount of visualization design candidates, it remains an open challenge on how to efficiently evaluate designs and select the best visualization designs catering to users' characteristics (e.g., preferences, perceptual, and cognitive capabilities) and contexts (e.g., tasks, devices, and environments).

This research concerns the online evaluation of visualization designs considering users' characteristics and contexts.
In human-computer interaction (HCI) as well as human factors and ergonomics (HFE) communities,
visualization designs have been evaluated from different perspectives by using various well-established instruments. Generally, qualitative evaluation methods have been proposed based on multivariate evaluation metrics that are collected by using questionnaires, think-aloud protocol, and interviews (\citealt{salvendy2012handbook}). However, these methods are typically time-consuming due to the data collection methods. Therefore, existing qualitative methods are not very suitable to support online evaluation of visualization designs.
Most of these methods also fail to quantify individual differences in users' characteristics and contexts. Although quantitative methods have been proposed to evaluate visualization designs by collecting objective data in an unobtrusive manner, most of them either fail to consider users' status (\citealt{ivory2001state}), or could only provide univariate response that provids limited insights for the investigators (\citealt{chen2017statistical}).

\begin{figure}
\begin{center}
\includegraphics[width=6.5in]{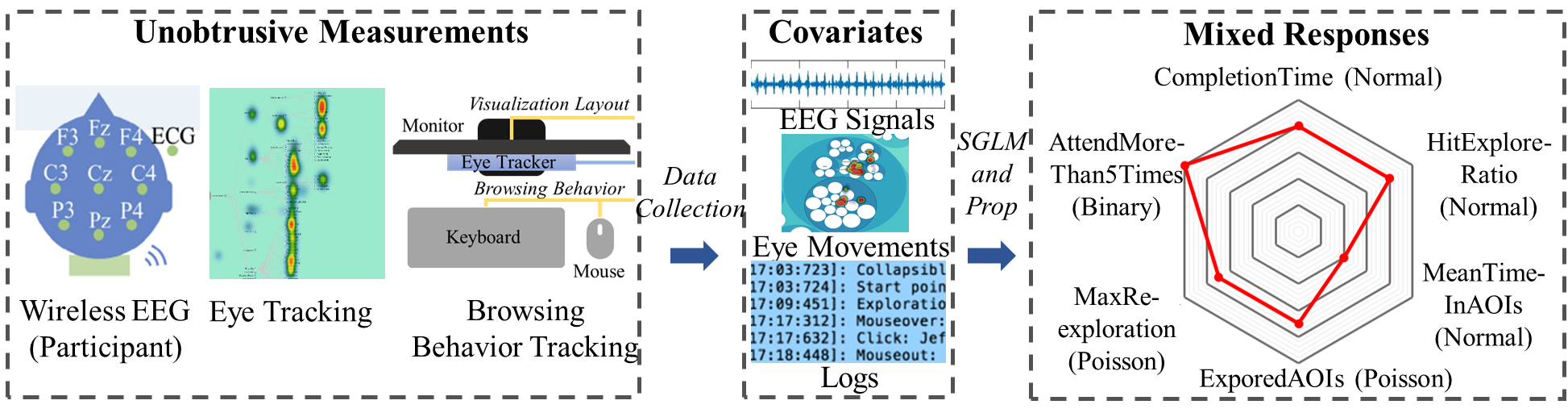}
\end{center}
\caption{A graphical illustration of data collection methods based on wireless EEG device, eye tracking device, and a browsing behavior logging system. Three types of covariates and six responses were extracted. (This figure is partially redrawn from \cite{chen2017statistical} with authors' permission) \label{fig:covariates_resps}}
\end{figure}

Obviously, a single response may not be as informative as multivariate responses to support in-depth evaluation of visualization designs from multiple perspectives.
As shown in the information visualization example in Figure~\ref{fig:covariates_resps}, multivariate responses are commonly collected with mixed types (i.e., continuous, counting, and binary responses). Specifically,
Figure \ref{fig:covariates_resps} shows a visualization evaluation user study by using unobtrusive data collection devices (i.e., wireless EEG device, remote eye tracking device, and a browsing behavior logging system).
After collecting the data, three types of measurements (i.e., EEG signals, eye movements, and behavioral logs) with six mixed responses were extracted, including three continuous responses (i.e., \textit{CompletionTime}, \textit{HitExploreRatio}, and \textit{MeanTimeInAOIs}), two counting responses (i.e., \textit{ExploredAOIs} and \textit{MaxRe-exploration}), and one binary response (i.e., \textit{AttendMoreThan5Times}) (see details in Section \ref{sec:app}).
To evaluate the visualization designs in a comprehensive manner, it calls for a multivariate regression for the mixed responses.
Such a data-driven model should be able to 1) predict mixed responses by considering the associations between themselves,
and 2) select significant predictor variables with meaningful interpretations.

In this work, we focus on developing a new multivariate regression for mixed responses (MRMR), which is to jointly model the mixed responses, and hence to improve the quantitative evaluation of the online visualization designs in accuracy.
The proposed method properly exploits the hidden associations among the multiple mixed responses, which draws a clear contrast to existing methods with separate model for each response.
Clearly, ignoring the dependency among multiple responses may lose information from data.
For the proposed MRMR method, we have considered a joint multivariate generalized linear model to fit data with mixed responses.
The key idea is to capture the associations between multiple responses via graphical model and improve the prediction performance via penalized regression.
It is thus able to select significant predictor variables such that the model is interpretable.
The Monte Carlo EM (MCEM) algorithm is adopted for efficient parameter estimation.

The rest of this paper is organized as follows. Section \ref{sec:literature} provides a literature review.
Section \ref{sec:model} introduces the proposed MRMR model and derives the objective function. The MCEM algorithm is developed to solve the proposed estimate in Section \ref{sec:MCEM}.
Section \ref{sec:sim} validates and stresses the proposed MRMR model by the numerical study.
An application of a visualization evaluation user study is thoroughly discussed in Section \ref{sec:app}.
We conclude this work with some discussion in Section \ref{sec:summary}.

\section{Literature Review}
\label{sec:literature}
In this section, we review the existing literature related to the qualitative and quantitative visualization design evaluation methods, as well as the mixed response modeling, respectively, in three subsections.

\subsection{Qualitative Visualization Design Evaluation}
\label{sec:qualitativeviseval}
Generally, existing methods for visualization design evaluation can be categorized into two groups: qualitative and quantitative evaluation methods.
Qualitative methods typically provide multivariate evaluation metrics which are subjectively rated by users and can be used for the generation of insightful conclusions for investigators. For example, testbed evaluation (\citealt{bowman1999testbed}) and sequential evaluation (\citealt{gabbard1999user}) methods were proposed for usability evaluation by using heuristics. The responses collected from user studies effectively identified usability issues against design principles from multiple perspectives (\citealt{bowman2002survey}).
As another example, pluralistic walkthrough (\citealt{bias1994pluralistic}) and cognitive walkthrough (\citealt{rieman1995usability}) evaluated visualization designs by collecting users' subjective responses in a series of pre-defined tasks according to their perceptual and cognitive workload.
Most of the aforementioned qualitative methods collect data based on well-designed questionnaires, think-aloud protocol, and interviews (\citealt{salvendy2012handbook}).
They typically take hours to collect data and days to extract informative data from records manually, thus being time-consuming to support visualization evaluation in a timely and online manner.
Besides these qualitative evaluation methods, various instruments such as NASA task load index (TLX) (\citealt{hart1988development}), subjective workload assessment technique (SWAT) (\citealt{reid1988subjective}), and software usability measurement inventory (SUMI) (\citealt{kirakowski1993sumi}) have been widely adopted in collecting the subjective feedback from user studies to provide multivariate evaluation scores.
Although these methods and instruments can support in-depth evaluation of visualization designs, they need a large amount of time for user studies and response collection, and they bring interference to the human-computer interactions.

\subsection{Quantitative Visualization Design Evaluation}
\label{sec:qualitativeviseval}
On the other hand, quantitative methods have been reported to evaluate visualization designs by collecting objective data in an unobtrusive manner.
For example, controlled experiments such as A/B testing were reported to evaluate the causal relationship between changes and their influence on measurable behaviors (\citealt{kohavi2009controlled}). Besides, an automated usability evaluating approach was proposed by using the background logging system to record relevant interaction information about users and visualization designs (\citealt{ivory2001state}).
These quantitative methods provide objective evaluations with minimal interference, but they fail to consider the users' characteristics, which may have significant impact on the evaluation conclusions and insights.
In contrast, some recent studies in HFE started to investigate data fusion and statistical models to predict subjective responses based on unobtrusive measures (e.g., physiological sensor signals, mouse and keyboard tracking, etc.).
For instance, \cite{chen2017statistical} employed a regularized linear regression model to accurately predict users' subjective rating of task complexities based on wearable sensor signals (i.e., electroencephalogram (EEG) signals, eye movements, and browsing behaviors).
Their method provided an online and quantitative evaluation model which yielded the first attempt to link unobtrusive measurements and subjective ratings.
As another example, \cite{maman2017data} built a regression model to predict subjective rating of physical fatigue by using wearable sensor signals (i.e., heart rate and vibrations of human bodies).
Note that the aforementioned data-driven models often consider a single response based upon wearable sensor signals (i.e., objective data) to predict subjective ratings. Therefore, it is relatively difficult for these data-driven methods to provide insights as informative as multivariate responses in qualitative methods.

\subsection{Mixed Response Modeling}
\label{sec:mixmdl}
In the literature, there are some works on joint modeling of multiple responses.
However, most related works limited their investigation on the joint modeling for mixed responses of only continuous and binary responses (\citealt{fitzmaurice1997regression, qiu2008distribution, song2009joint, chen2014selection, deng2015qq, kang2018bayesian}).
They often considered a conditional model of one-type variables conditioned on the other-type variables.
Besides, \cite{hapugoda2017joint} studied a set of poultry industry data with one continuous response and one counting response.
They formulated a joint model by combining the discrete time hazard model (\citealt{hapugoda2016joint}) and Poisson regression model.
The generalized linear models were used to form marginal models for each response.
Then the two responses were linked by structuring a covariance matrix to account for the potential correlations.
However, such a covariance matrix contains a user specified covariance structure.
Additionally, \cite{wu2018sparse} proposed to jointly model data with only multivariate counting responses by the latent variables.
The R pacakge SabreR (\citealt{berridge2011multivariate}) can model data with mixed responses by using the marginal likelihood but is limited to dealing with at most three response variables.
\cite{bonat2016multivariate} proposed a multivariate covariance generalized linear model to fit mixed responses and capture the relationship between response variables by the generalized Kronecker product.
However, their method needs to pre-specify a covariance link function and several known matrices to reflect the covariance structure of responses, and obviously such known matrices are subjectively defined by the users.
Therefore, based on our best knowledge, few works have contributed to the analysis of mixed responses data with continuous, counting and binary outcomes together.
A mixed response model is desired to consider the dependency among multiple responses with mixed types to support insightful online visualization evaluation.

\section{The Proposed Model with Exponential Family}
\label{sec:model}
In this section, we construct the proposed model and derive its likelihood function.
Suppose $\boldsymbol{U} = (U^{(1)}, U^{(2)}, \ldots, U^{(l)})^{T}$ is a random
vector of $l$ dimensions and each $U^{(i)}$ is a normal random variable
representing continuous response.
Similarly, $\boldsymbol{Z} = (Z^{(1)}, Z^{(2)}, \ldots, Z^{(m)})^{T}$ is a random vector of $m$ dimensions and each $Z^{(i)}$ is a Poisson random variable representing counting response, and
$\boldsymbol{W} = (W^{(1)}, W^{(2)}, \ldots, W^{(k)})^{T}$ is a random vector of
$k$ dimensions and each $W^{(i)}$ is a Bernoulli random variable representing binary response.
Then, we have
\begin{align}
	U^{(i)} | \mu^{(i)}, {\sigma^{(i)}} & \sim N\left(\mu^{(i)}, {\sigma^{(i)}}^{2}\right) \nonumber \\
	Z^{(i)} | \beta^{(i)} & \sim \text{Poisson}\left(\beta^{(i)}\right) \label{eq:model} \\
	W^{(i)} | \gamma^{(i)} & \sim \text{Bernoulli}\left(\gamma^{(i)}\right), \nonumber
\end{align}
where $\mu^{(i)}, {\sigma^{(i)}}^{2}, \beta^{(i)}$ and $\gamma^{(i)}$ are the corresponding distribution parameters.
Given such parameters, the model in \eqref{eq:model} implies that the variables $U^{(i)}$, $Z^{(i)}$ and $W^{(i)}$ are conditionally independent from each other.
Let $\boldsymbol{\xi}_G = (\mu^{(1)}, \ldots, \mu^{(l)})^T$,
$\boldsymbol{\xi}_{P} = (\log\beta^{(1)}, \ldots, \log\beta^{(m)})^T$ and
$\boldsymbol{\xi}_{B} = (\log\frac{\gamma^{(1)}}{1 - \gamma^{(1)}}, \ldots, \log\frac{\gamma^{(k)}}{1 - \gamma^{(k)}})^T$.
Consequently, we connect the vector of parameters $\boldsymbol{\xi} = (\boldsymbol{\xi}_{G}^T, \boldsymbol{\xi}_{P}^T, \boldsymbol{\xi}_{B}^T)^T$ with
the $p$-dimensional vector of predictor variables $\boldsymbol{x}$ via the linear model
\begin{align*}
	\boldsymbol{\xi} & = \boldsymbol{B}^T \boldsymbol{x} +
	\boldsymbol{\varepsilon} \\
	\boldsymbol{\varepsilon} & \sim N(\boldsymbol{0}, \boldsymbol{\Sigma}),
\end{align*}
where $\bm B$ is a $p \times (l + m + k)$ coefficient matrix,
and $\bm \Sigma$ is the covariance matrix of the error term $\bm \varepsilon$,
which characterizes the dependency between multiple responses.
From a Bayesian perspective, the parameter vector $\bm \xi$ is given a normal distribution with mean $\boldsymbol{B}^T \boldsymbol{x}$ and covariance matrix $\bm \Sigma$.
Denote the parameter vector by $\boldsymbol{\theta} = (\mu^{(1)}, \ldots, \mu^{(l)}, \beta^{(1)}, \ldots, \beta^{(m)}, \gamma^{(1)}, \ldots, \gamma^{(k)})^T$.
Accordingly, we have an element-wise mapping $\pi$ such that
$\boldsymbol{\theta} = \pi (\boldsymbol{\xi})$.
Specifically, $\pi$ is a mapping on an $(l + m +k)$ dimensional vector.
For the first $l$ components of the vector: $\xi \stackrel{\pi}{\rightarrow} \xi$;
for the next $m$ components of the vector: $\xi \stackrel{\pi}{\rightarrow} \exp{(\xi)}$;
for the last $k$ components of the vector: $\xi \stackrel{\pi}{\rightarrow} \frac{\exp{(\xi)}}{1 + \exp{(\xi)}}$.
Hence, we propose the following multivariate regression model
\begin{align}\label{eq:model2}
	\boldsymbol{Y} & \sim \text{Exponential Family}(\boldsymbol{\theta})  \nonumber \\
	\boldsymbol{\theta} & = \pi (\boldsymbol{B}^T \boldsymbol{x} +
	\boldsymbol{\varepsilon}) \\
	\boldsymbol{\varepsilon} & \sim N(\boldsymbol{0}, \boldsymbol{\Sigma}), \nonumber
\end{align}
where $\boldsymbol{Y} = [\boldsymbol{U}^T, \boldsymbol{Z}^T, \boldsymbol{W}^T]^T$ represents the vector of the response variables.
Through the parameter vector $\bm \theta$, we are able to model the covariance structure of the variable $\bm Y$ indirectly.

Before diving into the derivation of the likelihood of model \eqref{eq:model2},
we first examine the normal component in the regression model
\begin{align}\label{eq:submodel}
	\boldsymbol{U} & \sim N(\boldsymbol{\mu}, \boldsymbol{\Sigma}_{U}) \nonumber \\
	\boldsymbol{\mu} & = \boldsymbol{B}_{U}^T \boldsymbol{x} +
	\boldsymbol{\varepsilon} \\
	\boldsymbol{\varepsilon} & \sim N(\boldsymbol{0}, \boldsymbol{\Sigma}), \nonumber
\end{align}
where $\boldsymbol{B}_U$ and $\boldsymbol{\Sigma}_U$ are corresponding
coefficient and covariance matrices that are related to continuous responses.
Note that in this sub-model \eqref{eq:submodel},
the covariance structure $\bm \Sigma_{U}$ of the responses $\boldsymbol{U}$ is coupled with that of the error term $\bm \varepsilon$.
To avoid the model identifiability issue, we simplify the normal sub-model as follows
\begin{align*}
	\boldsymbol{U} & = \boldsymbol{\mu} \\
	\boldsymbol{\mu} & = \boldsymbol{B}_{U}^T \boldsymbol{x} +
	\boldsymbol{\varepsilon} \\
	\boldsymbol{\varepsilon} & \sim N(\boldsymbol{0}, \boldsymbol{\Sigma}_{U}).
\end{align*}
In this way, the covariance structure
of responses $\boldsymbol{U}$ can be modeled via the normally distributed error term
$\boldsymbol{\varepsilon}$, which reduces the number of parameters in the multivariate regression model.

With the simplification of the normal component in the regression model,
its likelihood function can be derived.
Under the conditional independence assumption that each dimension of $\boldsymbol{Y}$ is independent of each other given $\boldsymbol{\theta}$,
the joint distribution of $\boldsymbol{Y}$ is
\begin{align*}
	p(\boldsymbol{Y} \mid \boldsymbol{\theta}) &= \prod_{i=1}^{l}
	p(U^{(i)} = u^{(i)} \mid \boldsymbol{\theta}) \cdot \prod_{i=1}^{m} p(Z^{(i)} = z^{(i)} \mid
	\boldsymbol{\theta}) \cdot \prod_{i=1}^{k} p(W^{(i)} = w^{(i)} \mid \boldsymbol{\theta}) \\
    &= \prod_{i=1}^{m} \frac{(\beta^{(i)})^{z^{(i)}} \exp{(-\beta^{(i)})}}{z^{(i)} !} \cdot
    \prod_{i=1}^{k} (\gamma^{(i)})^{w^{(i)}} (1 - \gamma^{(i)})^{1 - w^{(i)}}. \nonumber
\end{align*}
From $\boldsymbol{\xi} =
\boldsymbol{B}^T \boldsymbol{x} + \boldsymbol{\varepsilon}$ and
$\boldsymbol{\varepsilon} \sim N(\boldsymbol{0}, \boldsymbol{\Sigma})$,
it is easy to derive the density function of $\boldsymbol{\xi} \mid \boldsymbol{x}$ as
\begin{align*}
	p(\boldsymbol{\xi} \mid \boldsymbol{x}) = \frac{1}{(2 \pi)^{(l + m +k)/2} |\boldsymbol{\Sigma}|^{1/2}} \exp{\left( -\frac{1}{2}(\boldsymbol{\xi} - \boldsymbol{B}^T \boldsymbol{x})^T \boldsymbol{\Sigma}^{-1} (\boldsymbol{\xi} - \boldsymbol{B}^T \boldsymbol{x}) \right)}.
\end{align*}
As a result, the density function of $\bm \theta \mid \boldsymbol{x}$ can be written as
\begin{align*}
p(\boldsymbol{\theta} \mid \boldsymbol{x}) = \frac{\exp{\left( -\frac{1}{2}[\pi^{-1} (\boldsymbol{\theta}) - \boldsymbol{B}^T \boldsymbol{x}]^T \boldsymbol{\Sigma}^{-1} [\pi^{-1} (\boldsymbol{\theta}) - \boldsymbol{B}^T \boldsymbol{x}] \right)}}{(2 \pi)^{(l + m +k)/2} |\boldsymbol{\Sigma}|^{1/2} \prod_{i=1}^m \beta^{(i)} \prod_{i=1}^k \gamma^{(i)} (1 - \gamma^{(i)})},
\end{align*}
where $\pi^{-1}$ is the inverse of mapping $\pi$.
Hence, the data distribution for $\boldsymbol{Y} \mid \boldsymbol{x}$ is
\begin{align}
	p(\boldsymbol{Y} \mid \boldsymbol{x}) = \int_{\boldsymbol{\theta}} p(\boldsymbol{Y},  \boldsymbol{\theta} \mid \boldsymbol{x}) d \boldsymbol{\theta} = \int_{\boldsymbol{\theta}}
	p(\boldsymbol{Y} \mid \boldsymbol{\theta}) p(\boldsymbol{\theta} \mid
	\boldsymbol{x}) d \boldsymbol{\theta}.
	\label{eq:data_distribution}
\end{align}
Given the training data $(\bm x_1, \bm y_1), (\bm x_2, \bm y_2), \ldots, (\bm x_n, \bm y_n)$, we write the log-likelihood of the regression model \eqref{eq:model2} as follows
\begin{align*}
	\mathcal{L}(\boldsymbol{B}, \boldsymbol{\Sigma}) = \sum_{j = 1}^{n} \log
	p(\boldsymbol{y}_j \mid \boldsymbol{x}_j).
\end{align*}

To jointly enforce the sparsity in the estimated model parameters, the LASSO-type regularization (\citealt{tibshirani1996regression}) is imposed on the matrices $\boldsymbol{B}$ and $\boldsymbol{\Sigma}$ simultaneously.
The penalty for $\bm B$ can recover the sparse structures on model coefficients and the penalty for $\bm \Sigma^{-1}$ is to enhance model prediction by incorporating the dependent relationship among the mixed responses.
Consequently, the regularized negative log-likelihood is
\begin{align*}
	\mathcal{L}_p(\boldsymbol{B}, \boldsymbol{\Sigma}) =
	-\mathcal{L}(\boldsymbol{B}, \boldsymbol{\Sigma}) +
	\lambda_1||\boldsymbol{B}||_1 + \lambda_2 ||\boldsymbol{\Sigma}^{-1}||_1,
\end{align*}
where $\lambda_1 > 0$ and $\lambda_2 > 0$ are two tuning parameters,
and $|| \cdot ||_{1}$ stands for the $L_1$ matrix norm, defined as $|| \bm A ||_{1} = \sum_{i,j} |a_{ij}|$ for matrix $\bm A$ with $a_{ij}$ being its elements.
Our proposed estimates $\hat{\bm B}$ and $\hat{\bm \Sigma}$ are the solution to the following optimization problem
\begin{align}\label{eq:optimization}
(\hat{\bm B}, \hat{\bm \Sigma}) = \mathop{\arg \min}\limits_{\bm B, \bm \Sigma}
\mathcal{L}_p(\boldsymbol{B}, \boldsymbol{\Sigma}).
\end{align}

\section{MCEM Algorithm for Parameter Estimation}
\label{sec:MCEM}
In this section, we apply the MCEM algorithm to handle the parameter estimation in \eqref{eq:optimization}.
The MCEM algorithm is a modified version of the EM algorithm where the expectation in the E-step is not available in a closed form. Alternatively, we approximate the expectation in the E-step by the numerical computation through Monte Carlo simulations.

\subsection{MCEM Algorithm}\label{subsec:MCEM}
In order to obtain the proposed estimates in \eqref{eq:optimization}, one needs to minimize the regularized negative log-likelihood function $\mathcal{L}_p(\boldsymbol{B}, \boldsymbol{\Sigma})$.
However, it is a difficult task due to the complicated integral in \eqref{eq:data_distribution}.
Therefore, we employ the EM algorithm for the parameter estimation with $\bm \theta$ treated as a latent variable.
The following gives an overview of EM algorithm applying to the optimization problem
\eqref{eq:optimization}, and the detailed procedure is provided in Section \ref{subsec:E} and  \ref{subsec:M}.

\textbf{E-step:} At iteration $t + 1$, the conditional distribution of $\bm \theta$ given
$\bm x$, $\bm Y$, $\bm B^{(t)}$ and $\bm \Sigma^{(t)}$ is
\begin{align*}
p(\bm \theta \mid \bm x, \bm Y, \bm B^{(t)}, \bm \Sigma^{(t)}) = \frac{p(\bm Y, \bm \theta \mid \bm x, \bm B^{(t)}, \bm \Sigma^{(t)})}{p(\bm Y \mid \bm x, \bm B^{(t)}, \bm \Sigma^{(t)})},
\end{align*}
where $\bm B^{(t)}$ and $\bm \Sigma^{(t)}$ are the estimates of $\bm B$ and $\bm \Sigma$
at iteration $t$ in the E-step. Then the expected log-likelihood is
\begin{align}\label{eq:Estep2}
Q(\bm B, \bm \Sigma \mid \bm B^{(t)}, \bm \Sigma^{(t)})
&= E_{\bm \theta \mid \bm x, \bm Y} [\mathcal{L}(\bm B, \bm \Sigma)]  \\
&= \sum_{j=1}^n E_{\bm \theta_j \mid \bm x_j, \bm y_j} [\log p(\bm y_j, \bm \theta_j \mid \bm x_j, \bm B^{(t)}, \bm \Sigma^{(t)})].   \nonumber
\end{align}

\textbf{M-step:} Find the estimates of $\bm B$ and $\bm \Sigma$ at iteration $t + 1$ by solving the following optimization problem
\begin{align*}
(\bm B^{(t+1)}, \bm \Sigma^{(t+1)}) = \mathop{\arg \min}\limits_{\bm B, \bm \Sigma} \{-Q(\bm B, \bm \Sigma \mid \bm B^{(t)}, \bm \Sigma^{(t)}) + \lambda_1||\boldsymbol{B}||_1 + \lambda_2 ||\boldsymbol{\Sigma}^{-1}||_1\}
\end{align*}
The E-step and M-step are repeated until the estimates of both $\bm B$ and $\bm \Sigma$ are converged.

\subsection{E-step}\label{subsec:E}
In this step, we take the expectation of the log-likelihood function as in \eqref{eq:Estep2}.
However, it is difficult to derive its analytical form due to the integral in \eqref{eq:data_distribution}.
Hence, the Markov Chain Monte Carlo (MCMC) technique is employed in the E-step to approximate the expected log-likelihood $Q(\bm B, \bm \Sigma \mid \bm B^{(t)}, \bm \Sigma^{(t)})$ in \eqref{eq:Estep2}.
That is, the MCMC samples of $\bm \theta_j$ are drawn from
\begin{align}\label{eq:MHobj}
&p(\bm \theta_j \mid \bm x_j, \bm y_j, \bm B^{(t)}, \bm \Sigma^{(t)}) \nonumber \\
\propto& p(\bm y_j, \bm \theta_j \mid \bm x_j, \bm B^{(t)}, \bm \Sigma^{(t)}) \nonumber \\
\propto& \exp{\left( -\frac{1}{2}[\pi^{-1} (\boldsymbol{\theta}) - \boldsymbol{B}^T \boldsymbol{x}]^T \boldsymbol{\Sigma}^{-1} [\pi^{-1} (\boldsymbol{\theta}) - \boldsymbol{B}^T \boldsymbol{x}] \right)} \nonumber \\
& \exp \left(-\sum_{i=1}^m \beta^{(i)} \right) \prod_{i=1}^m [ \beta^{(i)} ]^{(z^{(i)} - 1)} \prod_{i=1}^k [\gamma^{(i)}]^{(w^{(i)} - 1)} [1 - \gamma^{(i)}]^{-w^{(i)}}.
\end{align}
Then the expected log-likelihood function \eqref{eq:Estep2} is approximately computed by
\begin{align}\label{eq:approaQ}
\tilde{Q}(\bm B, \bm \Sigma \mid \bm B^{(t)}, \bm \Sigma^{(t)}) = \sum_{j=1}^n \frac{1}{h} \sum_{\nu=1}^h \log p(\bm y_j, \bm \theta_j^{(\nu)} \mid \bm x_j, \bm B^{(t)}, \bm \Sigma^{(t)}),
\end{align}
where $h$ is the size of MCMC samples after burn-in period.
In our implementation, we set the length of MCMC chain to be 1000 with the burn-in size 300.
The MCMC chain usually converges in one or two hundred iterations.

\subsection{M-step}\label{subsec:M}
In this step, we seek for the estimates of $\bm B$ and $\bm \Sigma$ to minimize the expected negative log-likelihood function (up to some constant)
\begin{align*}
\frac{1}{n} \sum_{j=1}^n \frac{1}{h} \sum_{\nu=1}^h \left\{ [\pi^{-1} (\bm \theta_j^{(\nu)}) - \bm B^T \bm x_j]^T \bm \Omega [\pi^{-1} (\bm \theta_j^{(\nu)}) - \bm B^T \bm x_j] - \log |\bm \Omega| \right\},
\end{align*}
where $\bm \Omega = \bm \Sigma^{-1}$. That is, we solve the following optimization problem
\begin{align}\label{eq:Mstep2}
(\bm B^{(t+1)}, \bm \Omega^{(t+1)}) = \mathop{\arg \min}\limits_{\bm B, \bm \Omega} \left\{ \frac{1}{nh} \mbox{tr} (\bm \Phi^T \bm \Phi \bm \Omega) - \log |\bm \Omega| + \lambda_1||\bm B||_1 + \lambda_2 ||\bm \Omega||_1 \right\},
\end{align}
where $\bm \Phi = \left (
\begin{array}{c}
\Psi (\bm \Theta_1) - \mathbb{X}_1 \bm B  \\
\Psi (\bm \Theta_2) - \mathbb{X}_2 \bm B  \\
\vdots  \\
\Psi (\bm \Theta_n) - \mathbb{X}_n \bm B  \\
\end{array}
\right)$, $\bm \Theta_j = (\bm \theta_{j}^{(1)}, \bm \theta_{j}^{(2)}, \ldots, \bm \theta_{j}^{(h)})^T$ is an $(n h) \times (l + m + k)$ matrix containing all the samples of $\bm \theta_j$ drawn from MCMC in E-step. Here
$\Psi (\bm \Theta_j) = [\pi^{-1}(\bm \theta_{j}^{(1)}), \pi^{-1}(\bm \theta_{j}^{(2)}), \ldots, \pi^{-1}(\bm \theta_{j}^{(h)})]^T$,
and $\mathbb{X}_j$ is an $h \times p$ matrix with each of its row being
$\bm x_j$. Notice that the optimization problem \eqref{eq:Mstep2} has two parameters $\bm B$ and $\bm \Omega$,
and it is convex with respect to one parameter when the other is fixed.
Hence, the profile technique is used to solve the optimization problem \eqref{eq:Mstep2}.
More precisely, for a known estimate $\bm B_0$,
\begin{align*}
\bm \Omega(\bm B_0) = \arg \min_{\bm \Omega} \left\{ \frac{1}{nh} \mbox{tr} (\bm \Phi^T \bm \Phi \bm \Omega) - \log |\bm \Omega| + \lambda_2 ||\bm \Omega||_1 \right\}.
\end{align*}
It has the same form as Graphical Lasso (Glasso), which has been studied in many literature such as \cite{yuan2007model}, \cite{friedman2008sparse}, \cite{rocha2008path}, \cite{rothman2008sparse}, \cite{raskutti2009model}, \cite{lam2009sparsistency}, \cite{deng2009large}, and \cite{yuan2010high}.
On the other hand, for a known estimate $\bm \Omega_0$,
\begin{align}\label{eq:solveB}
\bm B(\bm \Omega_0) = \arg \min_{\bm B} \left\{ \frac{1}{nh} \mbox{tr} (\bm \Phi^T \bm \Phi \bm \Omega_0) + \lambda_1 ||\bm B||_1 \right\}.
\end{align}
To solve the optimization problem \eqref{eq:solveB}, we approximate the $L_1$ matrix norm $||\bm B||_1$ by a quadratic form in order to reduce the computational burden.
Such a quadratic approximation technique is commonly used in the literature of sparse regressions, such as the SCAD penalty in \cite{fan2001variable} and \cite{zou2008one}.
Denote by $\hat{\bm B}$ the current estimate of $\bm B$.
Let $1 / \sqrt{|\hat{\bm B}|}$ be a matrix each of whose entries is the inverse of the squared root of the absolute value for the corresponding entry in matrix $\hat{\bm B}$. Then we have
\begin{align*}
\lambda_1 ||\bm B||_1 \approx \lambda_1 \mbox{tr} (\tilde{\bm B}^T \tilde{\bm B} ),~~~\mbox{with}~
\tilde{\bm B} = \bm B \circ \frac{1}{\sqrt{|\hat{\bm B}|}},
\end{align*}
where $\circ$ represents the Hadamard product (\citealt{Roger1985Matrix}).
Accordingly, the optimization problem \eqref{eq:solveB} can be written as
\begin{align*}
\eta(\bm B) &= \frac{1}{nh} \mbox{tr} (\bm \Phi^T \bm \Phi \bm \Omega_0) + \lambda_1 \mbox{tr} (\tilde{\bm B}^T \tilde{\bm B}) \\
&= \frac{1}{nh} \sum_{j=1}^n \mbox{tr} \left \{ [ \Psi(\bm \Theta_j) - \mathbb{X}_j \bm B]^T [ \Psi(\bm \Theta_j) - \mathbb{X}_j \bm B ] \bm \Omega_0 \right \} + \lambda_1 \mbox{tr} (\tilde{\bm B}^T \tilde{\bm B} ).
\end{align*}
Taking derivative of $\eta(\bm B)$ with respect to $\bm B$ and setting to $0$, we have
\begin{align}\label{eq:derB}
&\frac{2}{nh} \sum_{j=1}^n \left [ \mathbb{X}_j^T \mathbb{X}_j \bm B \bm \Omega_0 - \mathbb{X}_j^T \Psi(\bm \Theta_j) \bm \Omega_0 \right ] + \frac{2 \lambda_1}{|\hat{\bm B}|} \circ \bm B
= 0  \nonumber \\
&\left( \sum_{j=1}^n  \mathbb{X}_j^T \mathbb{X}_j \right) \bm B \bm \Omega_0 + \frac{\lambda_1 n h}{|\hat{\bm B}|} \circ \bm B = \left( \sum_{j=1}^n \mathbb{X}_j^T \Psi(\bm \Theta_j) \right) \bm \Omega_0.
\end{align}
Applying the matrix vectorization operator $\vectorize(\cdot)$ for both sides of \eqref{eq:derB} yields
\begin{align*}
\bm \Omega_0 \otimes \left( \sum_{j=1}^n \mathbb{X}_j^T \mathbb{X}_j \right) \vectorize(\bm B) + \vectorize(\frac{\lambda_1 n h}{|\hat{\bm B}|}) \circ \vectorize(\bm B) = \vectorize \left( \left[ \sum_{j=1}^n \mathbb{X}_j^T \Psi(\bm \Theta_j) \right] \bm \Omega_0 \right),
\end{align*}
where $\otimes$ represents the Kronecker product. As a result, we obtain the solution to the optimization problem \eqref{eq:solveB}
\begin{align*}
\vectorize(\bm B) = \left[ \bm \Omega_0 \otimes \left( \sum_{j=1}^n \mathbb{X}_j^T \mathbb{X}_j \right) + \mbox{diag} \left( \vectorize(\frac{\lambda_1 n h}{|\hat{\bm B}|}) \right) \right]^{-1} \vectorize \left( \left[ \sum_{j=1}^n \mathbb{X}_j^T \Psi(\bm \Theta_j) \right] \bm \Omega_0 \right).
\end{align*}
Based on the above discussion, the estimation procedure for the proposed model is thus summarized in Algorithm \ref{alg1} as follows:
\begin{algorithm} \label{alg1}
~

\textbf{Step 1}: Input initial values $\bm \Sigma_{init}$, $\bm B_{init}$, $\lambda_1$ and $\lambda_2$.

\textbf{Step 2}: Use the Metropolis-Hasting algorithm to draw samples of $\bm \theta_j$ from \eqref{eq:MHobj}.

\textbf{Step 3}: $\bm \Phi^{(t+1)} = \left (
\begin{array}{c}
\Psi (\bm \Theta_1) - \mathbb{X}_1 \bm B^{(t)}  \\
\Psi (\bm \Theta_2) - \mathbb{X}_2 \bm B^{(t)}  \\
\vdots  \\
\Psi (\bm \Theta_n) - \mathbb{X}_n \bm B^{(t)}  \\
\end{array}
\right)$.

\textbf{Step 4}: Obtain $\bm \Omega^{(t+1)}$ = Graphical Lasso $(\bm \Phi^{(t+1)}, \lambda_2)$.

\textbf{Step 5}: Obtain $\bm B^{(t+1)}$ as \\
$\vectorize(\bm B^{(t+1)}) = \left[ \bm \Omega^{(t+1)} \otimes \left( \sum_{j=1}^n \mathbb{X}_j^T \mathbb{X}_j \right) + \diag \left( \vectorize(\frac{\lambda_1 n h}{|\bm B^{(t)} |}) \right) \right]^{-1} \vectorize \left( \left[ \sum_{j=1}^n \mathbb{X}_j^T \Psi(\bm \Theta_j) \right] \bm \Omega^{(t+1)} \right)$.

\textbf{Step 6}: Repeat Step 2 - 5 till convergence.
\end{algorithm}

To choose the optimal values for the tuning parameters $\lambda_1$ and $\lambda_2$,
we use the extended Bayesian information criterion (EBIC) (\citealt{chen2008extended}) which is to balance the tradeoff between the fitting of the likelihood function and the sparsity of the estimates.
The EBIC is commonly used in the Gaussian graphical models with sparsity when dimensions are large.
Denote by $\hat{\bm B}_{\lambda_1, \lambda_2}$ and $\hat{\bm \Omega}_{\lambda_1, \lambda_2}$
the estimates of $\bm B$ and $\bm \Omega$ under the tuning parameters $(\lambda_1, \lambda_2)$.
Then the EBIC is computed as
\begin{align*}
\mbox{EBIC}(\lambda_1, \lambda_2) = &-2 \tilde{Q}(\hat{\bm B}_{\lambda_1, \lambda_2}, \hat{\bm \Omega}_{\lambda_1, \lambda_2}) + [\upsilon(\hat{\bm B}_{\lambda_1, \lambda_2}) + \upsilon(\hat{\bm \Omega}_{\lambda_1, \lambda_2})] \log n \\
 &+ 2 \tau \upsilon(\hat{\bm B}_{\lambda_1, \lambda_2}) \log[p(l+m+k)] + 4 \tau \upsilon(\hat{\bm \Omega}_{\lambda_1, \lambda_2}) \log(l+m+k),
\end{align*}
where $\tilde{Q}(\hat{\bm B}_{\lambda_1, \lambda_2}, \hat{\bm \Omega}_{\lambda_1, \lambda_2})$ can be approaximated by \eqref{eq:approaQ}.
Here $\upsilon(\hat{\bm B}_{\lambda_1, \lambda_2})$ and $\upsilon(\hat{\bm \Omega}_{\lambda_1, \lambda_2})$ represent the number of non-zeros in the estimates $\hat{\bm B}_{\lambda_1, \lambda_2}$ and $\hat{\bm \Omega}_{\lambda_1, \lambda_2}$, respectively.
$\tau$ is set to be 0.5 as suggested by \cite{foygel2010extended}.
The optimal values for ($\lambda_1$, $\lambda_2$) are chosen to minimize the EBIC.

\section{Numerical Simulation}\label{sec:sim}
In this section, the performance of the proposed method is examined by comparing with the separate models for each response without considering the dependency between response variables.
We use the following three structures for matrix $\bm \Omega$:

$\textbf{Example}$ 1. $\bm \Omega_{1} = \bm L^T \bm L$. The elements of matrix $\bm L$ are randomly generated from uniform distribution $Unif(-1, 1)$.

$\textbf{Example}$ 2. $\bm \Omega_{2}$ = MA(0.8, 0.6, 0.4, 0.2). The main diagonal elements are 1 with the $i$th sub-diagonal elements $0.2*(5-i), i = 1,2,3,4$.

$\textbf{Example}$ 3. $\bm \Omega_{3}$ is generated by randomly permuting rows and corresponding columns of $\bm \Omega_{2}$.

$\textbf{Example}$ 1 is a dense matrix, which can be considered as the most common matrix for the real-world data, such as social, financial and economic data.
$\textbf{Example}$ 2 has a banded sparse structure taking ones on its diagonal and decreasing values on several sub-diagonals with the rest elements being zeros.
It implies that the variables far apart are weakly correlated. This type of matrix usually occurs in longitudinal, spatial and time series data.
$\textbf{Example}$ 3 considers an unstructured sparse matrix, which is often the case for the high-dimensional data, such as gene expressions and image data.
For each example, two scenarios of dimensions are considered:
(1) $p = 30$, $l = 3$, $m = 3$ and $k = 3$;
(2) $p = 70$, $l = 5$, $m = 5$ and $k = 5$.
We independently generate $n = 50$ training observations and 30 testing observations of the predictor matrix from multivariate normal distribution $N_p(\bm 0, \sigma_X \bm I)$.
The true values of the coefficient matrix $\bm B$ are sampled from uniform distribution $Unif(a_B, b_B)$.
To enforce the sparsity, a proportion of zeros, denoted by $s_B$, is randomly placed into each column of matrix $\bm B$.
The observations in the response matrix are generated based on the models \eqref{eq:model} and \eqref{eq:model2}.
In order to investigate the impact of the variation in the predictor data on the performance of the proposed method, we scale the matrix $\bm \Sigma$ such that the largest element in $\bm \Sigma$ equals different values $\varphi = (1, 1.8, 2.6, 3.4)$.
By tuning the data generation parameters $\sigma_X$, $a_B$, $b_B$ and $s_B$,
we could make sure that the counting observations in the response matrix are within a reasonable range.

We compare the performance of the proposed method (Proposed) with the separate generalized linear model (SGLM), which models each response separately using LASSO-penalized generalized linear regression (\citealt{park2007l1, koh2007interior}).
Note that the SGLM ignores the associations between responses.
We implement the SGLM using \textit{glmnet} package in R program.
The Bayesian Information Criterion (BIC) is used to choose the optimal tuning parameters for SGLM.

\begin{table}[h]
\begin{center}
\caption{The averages and standard errors (in parenthesis) of loss measures for estimates.}
\label{table:est}
\resizebox{\textwidth}{!}{ 
\begin{tabular}{ccccccccccccccc}
\hline
  \multirow{3}*&&\multicolumn{4}{c}{$L(\hat{\bm B})$}&\multicolumn{2}{c}{$L(\hat{\bm \Omega})$} \\
\cline{3-8}
  &&\multicolumn{2}{c}{$p = 30$}&\multicolumn{2}{c}{$p = 70$}&\multicolumn{1}{c}{$p = 30$}&\multicolumn{1}{c}{$p = 70$}\\
\cline{3-8}
 &$\varphi$     &SGLM  &Proposed &SGLM  &Proposed  &Proposed  &Proposed \\
\hline
\multirow {4}*{$\bm \Omega_1$}
&1.0    &9.622 (1.244) &5.800 (0.047) &11.13 (0.209) &9.561 (0.076) &11.76 (0.075)  &27.31 (0.041)\\
&1.8    &18.42 (1.636) &5.652 (0.058) &13.81 (0.289) &9.458 (0.065) &11.83 (0.105)  &24.02 (0.041)\\
&2.6    &35.40 (2.104) &5.590 (0.085) &17.62 (0.781) &9.570 (0.054) &10.58 (0.058)  &21.63 (0.044)\\
&3.4    &47.09 (1.819) &5.917 (0.055) &21.43 (0.632) &9.700 (0.070) &12.76 (0.041)  &27.33 (0.173)\\
\midrule
\multirow {4}*{$\bm \Omega_2$}
&1.0    &10.84 (0.537) &5.736 (0.080) &11.21 (0.161) &9.508 (0.070) &5.503 (0.104)  &8.012 (0.008)\\
&1.8    &13.50 (0.924) &5.690 (0.057) &13.56 (0.162) &9.446 (0.070) &5.299 (0.093)  &8.035 (0.004)\\
&2.6    &18.67 (1.549) &5.720 (0.049) &16.56 (0.277) &9.649 (0.058) &5.195 (0.092)  &8.043 (0.004)\\
&3.4    &35.64 (2.283) &5.968 (0.059) &22.02 (0.463) &9.709 (0.075) &5.422 (0.100)  &8.031 (0.007)\\
\midrule
\multirow {4}*{$\bm \Omega_3$}
&1.0    &9.511 (0.849) &5.823 (0.113) &11.07 (0.145) &9.397 (0.062) &5.332 (0.104) &8.061 (0.007)\\
&1.8    &14.07 (1.285) &5.780 (0.060) &13.45 (0.160) &9.402 (0.069) &5.375 (0.109)  &8.050 (0.005)\\
&2.6    &23.50 (2.070) &5.796 (0.047) &17.62 (0.286) &9.596 (0.076) &5.447 (0.107)  &8.071 (0.005)\\
&3.4    &32.50 (2.536) &5.930 (0.094) &21.81 (0.479) &9.771 (0.065) &5.681 (0.120)  &8.067 (0.008)\\
\midrule
\hline
\end{tabular}}
\end{center}
\end{table}

To evaluate the estimation accuracy of the proposed method with respect to the coefficient matrix $\bm B$ and precision matrix $\bm \Omega$, we consider the loss measures as follows
\begin{align*}
L(\hat{\bm B}) = || \bm B - \hat{\bm B} ||_F^2 ~~~ \mbox{and} ~~~ L(\hat{\bm \Omega}) = || \bm \Omega - \hat{\bm \Omega} ||_F^2,
\end{align*}
where $||\cdot||_F$ represents the Frobenius norm.
Here, $\hat{\bm B}$ and $\hat{\bm \Omega}$ stand for estimates of $\bm B$ and $\bm \Omega$, respectively.
Table \ref{table:est} reports the averages and corresponding standard errors of the loss measures for the estimates $\hat{\bm B}$ and $\hat{\bm \Omega}$ over 50 replicates.
It is clear to see that the proposed method consistently outperforms the SGLM in all settings.
As the variation in the simulated data increases ($\varphi$ increases), the traditional method SGLM produces large loss measures regarding $L(\hat{\bm B})$, while the proposed method shows a more stable performance.
In addition, the proposed model is designated to be able to capture the sparse structure in the underlying matrix.
Thus, it can provide more accurate estimation if the underlying matrix is sparse, which is evidenced by the results of $L(\hat{\bm \Omega})$ that the losses for $\bm \Omega_2$ and  $\bm \Omega_3$ are relatively smaller than those for $\bm \Omega_1$.
Such results demonstrate that the proposed MRMR model improves the estimation accuracy by modeling the dependency structure between the multivariate mixed responses.

\begin{table}[h]
\begin{center}
\caption{The averages and standard errors (in parenthesis) of RMSE and ME when $p = 30$.}
\label{table:RMSEp30}
\resizebox{\textwidth}{!}{ 
\begin{tabular}{ccccccccccccccc}
\hline
  \multirow{2}*&&\multicolumn{2}{c}{$L$(Norm)}&&\multicolumn{2}{c}{$L$(Poisson)}&&\multicolumn{2}{c}{$L$(Binary)} \\
\cline{3-10}
 &$\varphi$     &SGLM  &Proposed &&SGLM  &Proposed  &&SGLM  &Proposed  \\
\hline
\multirow {4}*{$\bm \Omega_1$}
&1.0    &1.393 (0.020) &1.499 (0.017) &&22.01 (3.275) &10.25 (1.303) &&0.321 (0.004) &0.263 (0.004) \\
&1.8    &1.664 (0.021) &1.754 (0.019) &&75.21 (6.590) &11.42 (1.199) &&0.325 (0.006) &0.265 (0.005) \\
&2.6    &1.977 (0.035) &1.951 (0.024) &&178.2 (37.13) &17.02 (1.634) &&0.325 (0.004) &0.265 (0.005) \\
&3.4    &2.293 (0.042) &2.111 (0.050) &&633.9 (176.9) &25.10 (1.650) &&0.326 (0.005) &0.266 (0.007) \\
\midrule
\multirow {4}*{$\bm \Omega_2$}
&1.0    &1.362 (0.018) &1.503 (0.017) &&18.13 (2.943) &8.540 (0.720) &&0.317 (0.006) &0.260 (0.005)\\
&1.8    &1.640 (0.020) &1.726 (0.021) &&122.1 (34.63) &14.47 (1.553) &&0.322 (0.005) &0.263 (0.004)\\
&2.6    &2.031 (0.035) &1.958 (0.025) &&219.7 (61.37) &17.81 (1.499) &&0.321 (0.005) &0.262 (0.004)\\
&3.4    &2.356 (0.037) &2.142 (0.024) &&844.9 (195.4)  &25.20 (1.695) &&0.323 (0.005) &0.270 (0.004)\\
\midrule
\multirow {4}*{$\bm \Omega_3$}
&1.0    &1.360 (0.019) &1.488 (0.020) &&15.89 (3.713) &7.833 (0.831) &&0.314 (0.004) &0.259 (0.004)\\
&1.8    &1.674 (0.022) &1.728 (0.021) &&138.9 (32.26) &12.54 (1.160) &&0.302 (0.004) &0.259 (0.004)\\
&2.6    &1.979 (0.030) &1.936 (0.021) &&248.7 (52.80) &19.90 (1.537) &&0.316 (0.005) &0.264 (0.004)\\
&3.4    &2.303 (0.040) &2.132 (0.023) &&779.7 (132.9)  &23.94 (1.946) &&0.313 (0.005) &0.267 (0.004)\\
\midrule
\hline
\end{tabular}}
\end{center}
\end{table}

\begin{table}[h]
\begin{center}
\caption{The averages and standard errors (in parenthesis) of RMSE and ME when $p = 70$.}
\label{table:RMSEp70}
\resizebox{\textwidth}{!}{ 
\begin{tabular}{ccccccccccccccc}
\hline
  \multirow{2}*&&\multicolumn{2}{c}{$L$(Norm)}&&\multicolumn{2}{c}{$L$(Poisson)}&&\multicolumn{2}{c}{$L$(Binary)} \\
\cline{3-10}
 &$\varphi$     &SGLM  &Proposed &&SGLM  &Proposed  &&SGLM  &Proposed  \\
\hline
\multirow {4}*{$\bm \Omega_1$}
&1.0    &1.448 (0.017) &1.511 (0.017) &&9.886 (0.778) &8.270 (0.661) &&0.387 (0.004) &0.343 (0.004) \\
&1.8    &1.696 (0.019) &1.721 (0.017) &&18.06 (1.740) &13.61 (1.192) &&0.382 (0.005) &0.353 (0.006) \\
&2.6    &2.133 (0.047) &1.962 (0.024) &&28.30 (2.625) &19.81 (1.676) &&0.395 (0.005) &0.367 (0.005) \\
&3.4    &2.500 (0.058) &2.136 (0.036) &&55.09 (3.344) &28.04 (1.740) &&0.399 (0.006) &0.357 (0.004) \\
\midrule
\multirow {4}*{$\bm \Omega_2$}
&1.0    &1.452 (0.016) &1.520 (0.016) &&11.62 (0.731) &8.026 (0.698) &&0.372 (0.004) &0.333 (0.004)\\
&1.8    &1.743 (0.019) &1.748 (0.018) &&20.70 (1.584) &14.10 (0.942) &&0.372 (0.006) &0.332 (0.004)\\
&2.6    &2.074 (0.028) &1.972 (0.024) &&28.99 (2.338) &19.44 (1.110) &&0.366 (0.004) &0.336 (0.004)\\
&3.4    &2.536 (0.043) &2.164 (0.022) &&41.92 (3.230) &26.26 (1.595) &&0.383 (0.004) &0.340 (0.003)\\
\midrule
\multirow {4}*{$\bm \Omega_3$}
&1.0    &1.408 (0.012) &1.467 (0.013) &&13.25 (0.742) &7.565 (0.527) &&0.370 (0.005) &0.330 (0.005)\\
&1.8    &1.726 (0.016) &1.736 (0.016) &&18.76 (2.338) &13.22 (1.057) &&0.361 (0.004) &0.325 (0.004)\\
&2.6    &2.154 (0.029) &1.985 (0.019) &&27.42 (2.224) &18.71 (1.306) &&0.379 (0.005) &0.341 (0.004)\\
&3.4    &2.520 (0.032) &2.161 (0.016) &&42.33 (3.170) &25.03 (1.228) &&0.378 (0.004) &0.337 (0.004)\\
\midrule
\hline
\end{tabular}}
\end{center}
\end{table}

To further investigate the performance of the proposed method, we evaluate its prediction accuracy using the root-mean-square error (RMSE) for the continuous and counting responses over the testing data.
With respect to the binary response, we compare the misclassification errors (ME) produced by two models on the testing data.
Let $L$(Norm) and  $L$(Poisson) denote the RMSE for the estimates of continuous responses and counting responses.
Let $L$(Binary) be the ME for the estimates of binary responses.
The cut-off point for the binary response estimates is 0.5.
Tables \ref{table:RMSEp30} and \ref{table:RMSEp70} display the averages and corresponding standard errors of RMSE and ME for settings $p = 30$ and $p = 70$ over 50 replicates.
For the continuous responses, although the proposed method is slightly inferior to the SGLM when the variation in the data is small ($\varphi = 1, 1.8$), it has better performance when the variation becomes larger ($\varphi = 2.6, 3.4$).
Under $L$(Binary), the prediction performance of the proposed method is consistently superior over that of the SGLM.
As for the RMSE of counting responses, the proposed method substantially outperforms the SGLM, especially when $\varphi$ is large.
These results demonstrate that the proposed MRMR model improves the prediction performance by incorporating the dependent relationship between the mixed responses.
Besides, the proposed method is more robust than the SGLM regarding the variation in the data.
Here, we also notice that the SGLM for counting responses has numeric convergence issues sometimes in the setting of $p = 30$, especially when $\varphi$ is large, which is also observed in \cite{wu2018sparse}.
This explains the large values of $L$(Poisson) for SGLM in Table \ref{table:RMSEp30}.

\begin{figure}[h]
\centering
\subfloat[] {\label{MRMR:traceplot} \scalebox{0.4}[0.45]{\includegraphics{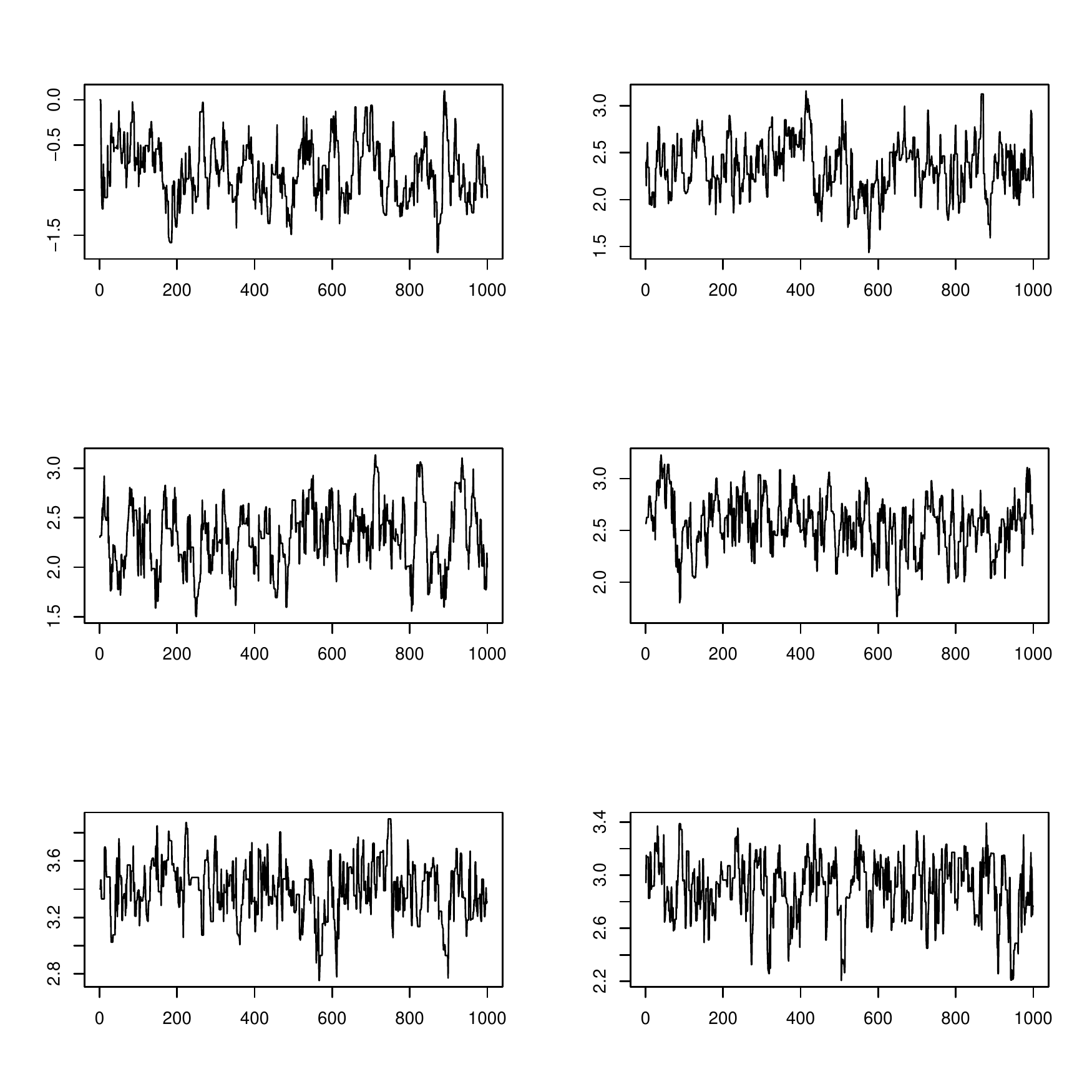}}}\hspace{0.1in}
\subfloat[] {\label{MRMR:ACF}
\scalebox{0.4}[0.45]{\includegraphics{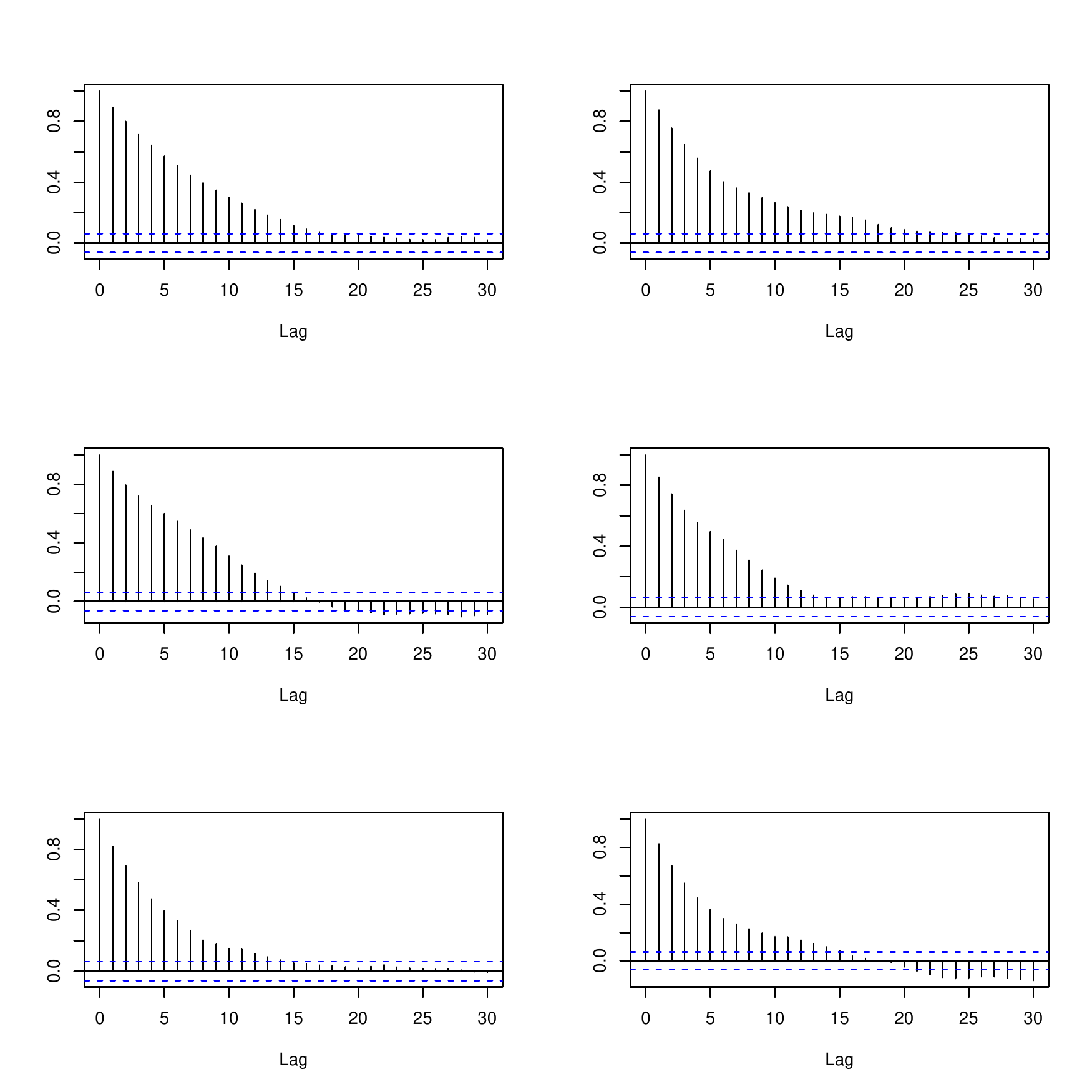}}}
\caption{Trace plots and ACF plots for the randomly selected parameters of one replicate from $\textbf{Example}$ 1 when $p = 30$ and $\varphi = 1$. (a) trace plots; (b) ACF plots.}
\label{fig:traceACF}
\end{figure}

In the proposed model, the Metropolis-Hasting method is employed to draw samples of $\bm \theta_j$ from \eqref{eq:MHobj} in the E-step of MCEM algorithm.
Note that the length of MCMC chain and the choice of burn-in size need to be pre-specified, while there is no general rule on the number of MCMC samples and burn-in period.
In our numerical examples, we often observe that the MCMC chain would converge in one or two hundred iterations.
Considering the computational burden, we suggest 1000 MCMC draws with the first 300 samples as burn-in period for the proposed model.
Figure \ref{fig:traceACF} displays the trace plots and corresponding autocorrelation function (ACF) plots for six randomly chosen parameters from one replicate of $\textbf{Example}$ 1 with $p = 30$ and $\varphi = 1$,
The results from the figure indicate that the 1000 MCMC samples and 300 burn-in size appear to be reasonable choices.
The rest of parameters have the similar patterns of trace plots and ACF plots, and hence omitted here.

\begin{table}[h]
\begin{center}
\caption{The averages and standard errors (in parenthesis) of loss measures when $p = 5$.}
\label{table:p5}
\begin{tabular}{ccccccccccccccccc}
\hline
\multicolumn{2}{c}{}&$\varphi = 1$ &$\varphi = 1.8$ &$\varphi = 2.6$ &$\varphi = 3.4$  \\
\hline
\multirow{3}{*}{$L(\hat{\bm B})$}
&SGLM       &1.052(0.061)     &1.505(0.034) &1.808(0.050) &2.241(0.073)\\
&MCGLM      &0.858(0.022)     &1.372(0.031) &1.675(0.050) &2.194(0.053)\\
&Proposed   &0.478(0.012)     &0.477(0.010) &0.485(0.010) &0.509(0.011)\\
\hline
\multirow{2}{*}{$L(\hat{\bm \Omega})$}
&MCGLM      &6.008(0.017)     &4.559(0.013) &8.233(0.016) &5.660(0.117)\\
&Proposed   &5.033(0.026)     &3.411(0.014) &6.501(0.022) &4.697(0.032)\\
\hline
\multirow{3}{*}{$L$(Norm)}
&SGLM       &1.051(0.018)     &1.418(0.021) &1.728(0.033) &1.885(0.049)\\
&MCGLM      &1.043(0.017)     &1.397(0.021) &1.681(0.031) &1.877(0.032)\\
&Proposed   &1.045(0.018)     &1.398(0.021) &1.649(0.028) &1.826(0.029)\\
\hline
\multirow{3}{*}{$L$(Poisson)}
&SGLM       &2.981(0.342)     &6.202(0.449) &10.48(1.096) &19.53(2.154)\\
&MCGLM      &2.649(0.113)     &6.087(0.433) &9.779(0.948) &17.78(1.612)\\
&Proposed   &2.666(0.121)     &5.858(0.419) &8.846(0.785) &14.21(1.328)\\
\hline
\multirow{3}{*}{$L$(Binary)}
&SGLM       &0.264(0.009)     &0.282(0.007) &0.289(0.006) &0.292(0.008)\\
&MCGLM      &0.243(0.007)     &0.257(0.006) &0.267(0.006) &0.272(0.006)\\
&Proposed   &0.237(0.007)     &0.241(0.006) &0.243(0.006) &0.251(0.006)\\
\hline
\end{tabular}
\end{center}
\end{table}

To further enhance the scope of the proposed method,
we conduct a simulation setting with the comparison method of multivariate covariance generalized linear model (MCGLM) by \cite{bonat2016multivariate}, which is implemented using R package \textit{mcglm}.
The MCGLM takes into account the possible association between the multivariate mixed responses when fitting them within the framework of generalized linear models.
It can provide an estimate of inverse covariance matrix characterizing the relationship between responses.
However, the MCGLM needs selecting a proper covariance link function and several known matrices before data analysis,
which reflects the association structure of responses.
The choice of a proper covariance link function is often subjectively made based on users's experience.
Additionally, the MCGLM encounters the algorithm convergence issue for large $p$. A possible explanation is that their algorithm does not incorporate the regularization technique.
In contrast, the proposed model takes advantage of the well-developed techniques of Lasso and Glasso to circumvent the computational convergence for high-dimensional data.
We thus run a numerical simulation on a small sale of $p = 5$, $l = 2$, $m = 2$ and $k = 1$ for $\textbf{Example}$ 1.
The generation of training set, testing set, the true coefficient matrix $\bm B$, the covariance matrix $\bm \Sigma$, the values of $\varphi$ and other relevant parameters remains the same as above.
Table \ref{table:p5} reports the averages and corresponding standard errors of all the loss measures of $L(\hat{\bm B})$, $L(\hat{\bm \Omega})$, $L$(Norm), $L$(Poisson) and $L$(Binary) for three methods over 50 replicates.
Clearly, it is seen that the proposed model is overall the best compared with the SGLM and MCGLM approaches.
It dominates the other two methods in terms of $L(\hat{\bm B})$ and produces lower values of $L(\hat{\bm \Omega})$.
Although the proposed model is comparable to the MCGLM regarding $L$(Norm), $L$(Poisson) and $L$(Binary) when $\varphi$ = 1 and 1.8,
it performs better as the value of variation parameter $\varphi$ increases.
The SGLM, overlooking the dependency between response variables, is inferior to the MCGLM and the proposed model.

\section{Case Study}\label{sec:app}

\begin{figure}
\begin{center}
\includegraphics[width=6.5in]{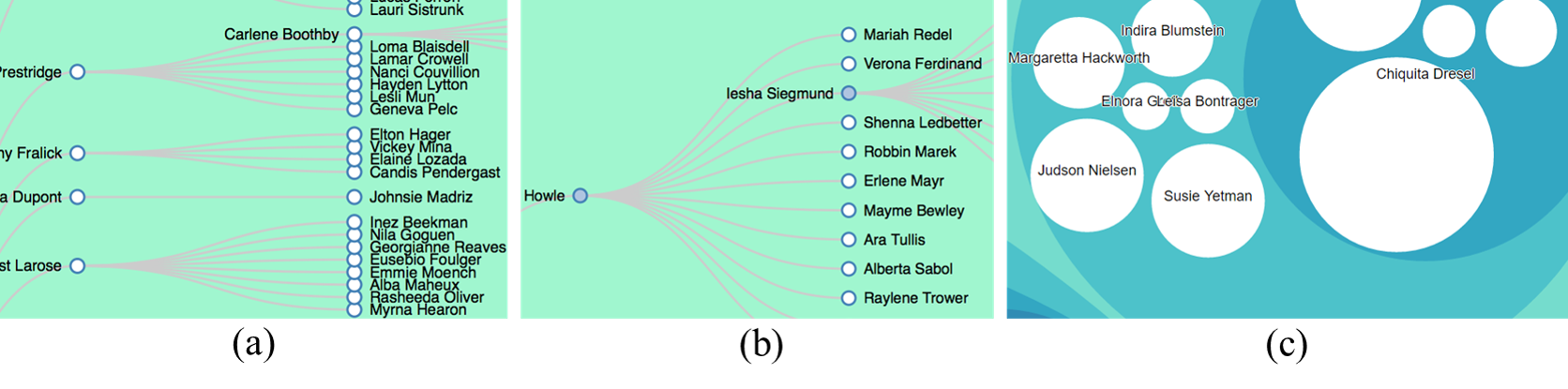}
\end{center}
\caption{Example of three visualization designs for user study, where (a) is the static node-link tree to map names and hierarchical relationships to the circles, texts, and edges, respectively; (b) is the collapsible node-link tree, which is an interactive variant of (a) with collapsible branches; and (c) is the zoomable pack layout to map the names and hierarchical relationships to nested and zommable circles (Redrawn from \cite{chen2017statistical} with authors' permission). \label{fig:vis_designs}}
\end{figure}

In this section, we demonstrate the merits of the proposed method through the case study of information visualization described in Section \ref{sec:intro}.
This data set was collected from a user study to quantitatively evaluate three interactive visualization designs (\citealt{chen2017statistical}), namely, the static node-link tree diagram, the collapsible node-link tree, and the zoomable layout (see Figure~\ref{fig:vis_designs}).
In this user study, 15 participants were recruited to evaluate three interactive visualization designs by performing 11 pre-defined visual searching tasks for each design. Note that data for one participant was removed due to missing values.

During performing tasks, as presented in Figure~\ref{fig:covariates_resps},
all participants' electroencephalogram (EEG) signals, eye movements, and browsing behaviors were collected by using an ABM\textregistered{} B-Alert 10-channel wireless EEG device, a SMI\textregistered{} REDn remote eye tracker, and a background logging system, respectively.
Following the feature extraction methods described in \cite{chen2017statistical}, 481 predictor variables were extracted from these sensor signals and browsing behaviors.
Some exploratory data analysis has been conducted to screen out the unimportant predictors, which resulted in 58 predictor variables for use.

In total, the data set contains $n= 462$ $(14\times11\times3)$ observations with three continuous response variables ($Y_{1}, Y_{2}, Y_{3}$), two counting response variables ($Y_{4}, Y_{5}$), and one binary response variable ($Y_{6}$).
The $Y_1$ is CompletionTime, which identifies the total time consumption for one participant to perform a task.
The $Y_2$ is HitExploreRatio,  a ratio of number of visually reached targets over number of explored area of interests (AOIs).
Here AOI is referred as a node in a node-link tree diagram, which associates with eye fixation for more than 0.5 second.
The $Y_3$ is MeanTimeInAOIs as average time spent in AOIs, which can be used to measure the level of confusion the AOIs caused for participants.
The counting response $Y_4$ is ExploredAOIs,  which counts the total number of explored AOIs in performing a task; and the $Y_5$ is MaxReexploration as the highest number of re-explorations at one AOI in performing a task.
The binary response is $Y_{6}$, AttendMoreThan5Times, which has value 1 when MaxReexploration is higher than five, and 0 otherwise.
These six response variables are to represent the efficiency of the visualization design from different perspectives.

We compare the performances of the proposed method with SGLM by considering two manners of data partition for training and testing.
The first manner is random splits: the whole data set is randomly partitioned into a training set with sample size 200 and a testing set with the rest 262 observations;
The second manner is leave-one-participant-out cross-validation (CV): one (33 observations) out of 14 participants is iteratively left out for testing, and the rest are used for training (429 observations).
Note that the leave-one-participant-out CV is investigated to stress the proposed model, since individual differences may easily lead to the violation of assumptions for linear model in a usability test that evaluates a user interface (\citealt{nielsen1990heuristic}).
The training set is used to fit the proposed model and SGLM, and the loss measures $L$(Norm), $L$(Poisson) and $L$(Binary) are computed from the testing data.
Table \ref{table:app} summarizes the averages of loss measures and corresponding standard errors from the random splits of 50 times, and the leave-one-participant-out CV.

\begin{table}
\begin{center}
\caption{The averages and standard errors (in parenthesis) of loss measures from 50 random splits of data and leave-one-participant-out cross-validation.}\label{table:app}
\begin{tabular}{ccccccccccccc}
\hline\hline
&   &               &$L$(Norm)      &$L$(Poisson)   &$L$(Binary)        \\\hline
&\multirow{2}{*}{Random Splits} &SGLM           &3.001 (0.015)  &115.3 (21.89)  &0.189 (0.003)      \\
&   &Proposed           &2.808 (0.019)  &36.72 (0.311)  &0.183 (0.004)      \\\hline
&\multirow{2}{*}{Leave-one-participant-out} &SGLM           &2.966 (0.121)  &47.82 (11.97)  &0.186 (0.014)      \\
&   &Proposed           &2.769 (0.146)  &36.38 (2.098)  &0.178 (0.016)      \\
\hline\hline
\end{tabular}
\end{center}
\end{table}

\begin{figure}[h]
\begin{center}
\scalebox{0.5}[0.5]{\includegraphics{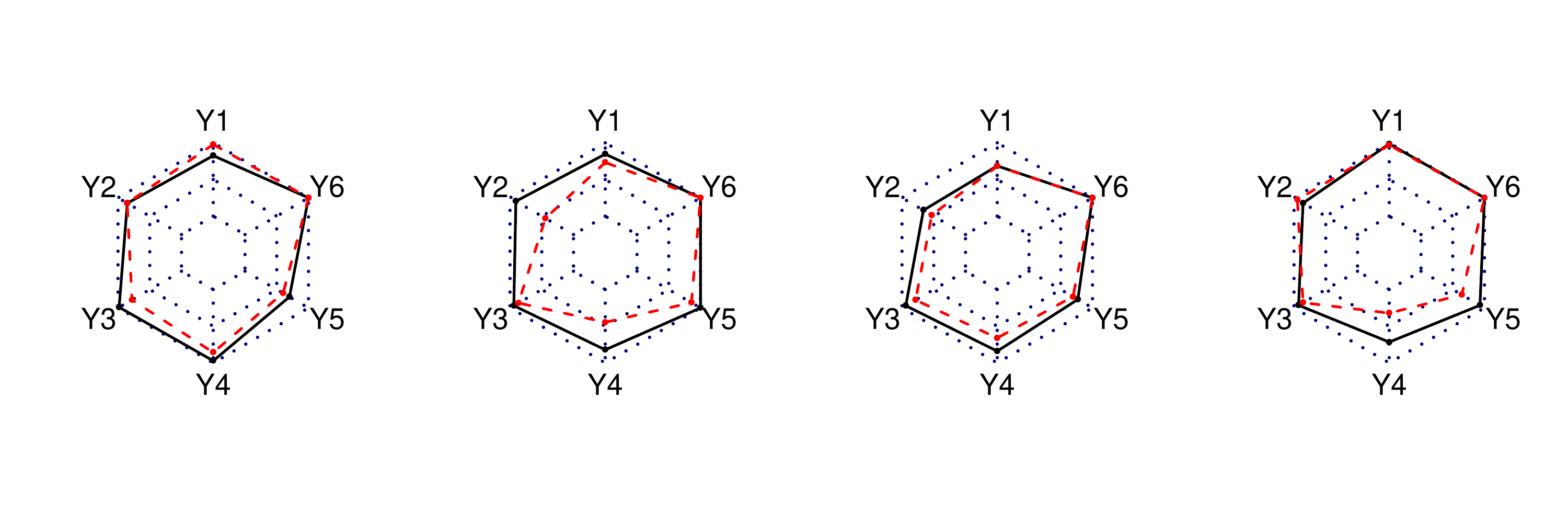}}
\caption{Radar maps for the predictions of the proposed MRMR and SGLM from randomly selected four samples of a testing data.
The hexagons represent the testing samples; the black solid lines are the predictions of the proposed MRMR; the red dashed lines indicate the predictions of the SGLM.} \label{fig:radar}
\end{center}
\end{figure}

\begin{table}
\begin{center}
\caption{The estimated correlation matrix of responses obtained by the proposed model  from one random split of data.}\label{table:omega}
{ 
\begin{tabular}{cccccccccccc}
\hline
&           &$Y_1$ &$Y_2$ & $Y_3$ & $Y_4$ & $Y_5$  &$Y_6$\\\hline
&$Y_1$   &1        &          &          &           &            &      \\
&$Y_2$   &-0.193   &1         &          &           &            &      \\
&$Y_3$   &0.060    &0.143     &1         &           &            &      \\
&$Y_4$   &0.036    &-0.011    &-0.046    &1          &            &      \\
&$Y_5$   &0.252    &-0.085    &0.009     &0.256      &1           &      \\
&$Y_6$   &0.670    &-0.224    &0.027     &0.119      &0.397       &1      \\
\hline
\end{tabular}}
\end{center}
\end{table}

From the table, it is seen that the proposed model generally outperforms the SGLM in terms of each loss measure for both random splits and leave-one-participant-out CV.
In particular, the proposed model largely improves the prediction accuracy for the counting responses with much smaller prediction error and standard error compared with the SGLM.
The proposed model is also superior in the prediction to the SGLM with respect to the continuous responses, and slightly better regarding the binary response.
These results demonstrate the advantage of the proposed MRMR model over the SGLM by considering the dependency relation between multiple responses.
In addition, to further evaluate the prediction ability with respect to the overall mixed responses,
Figure \ref{fig:radar} displays four radar maps corresponding to four testing data points randomly selected from one split of data.
The hexagons, solid lines and dashed lines stand for the testing data values, the prediction values from the proposed MRMR and the prediction values from the SGLM, respectively.
Hence, the closeness of the graphs composed of the solid or dashed lines to the hexagons illustrates the prediction ability of the corresponding models.
From Figure \ref{fig:radar}, it can be seen that although the proposed MRMR model with solid lines may be inferior sometimes with certain response ($Y_1$ for the first panel, and $Y_2$ for the fourth panel),
it generally performs better than the SGLM.
Furthermore, another advantage is that the proposed method is able to provide the estimate of correlation matrix of response variables, which captures the dependency relationship between responses.
Table \ref{table:omega} reports such an estimate from one random split of data.
Although there are some variables having weak correlations,
several relatively strong correlations exist in variables $Y_1$ and $Y_6$, $Y_5$ and $Y_6$, $Y_4$ and $Y_5$, $Y_1$ and $Y_5$, which cannot be ignored.
Besides, we also check each estimate of $\bm \Omega$ obtained from 50 splits of data and find that most estimates are dense matrices.
The mean of the largest absolute value in $\bm \Sigma$ is 1.72 with its standard deviation being 0.044.
Hence, the underlying matrix $\bm \Omega$ of the data might be closely represented by the $\textbf{Example}$ 1 with variation $\varphi = 1.8$ in Section \ref{sec:sim}.

In addition to the statistical performance, several interesting findings from a human-computer interaction (HCI) perspective can be identified from the superior performance of the proposed MRMR model.
Firstly, the proposed model provides accurate prediction for subjective responses in a visualization evaluation user study merely based on objective and unobtrusive measurements and a logging system.
In literature, interviews and questionnaires are typically adopted to provide subjective responses regarding the effectiveness and efficiency of the visualization designs (\citealt{bowman2002survey}).
But they typically lead to interference in participants' subjective responses and require huge amount of time to conduct the interviews and questionnaires.
As an enhancement to \cite{chen2017statistical}, which firstly proved the feasibility for quantitative and unobtrusive evaluation,
the proposed MRMR method enables the joint modeling of multiple mixed responses, providing more realistic techniques to support efficient evaluation studies in HCI.

Secondly, the results from leave-one-participant-out CV indicate that the proposed model can better address the individual differences in the user study since it provides better prediction performance for all three types of responses.
Such an advantage can be explained by the quantification of correlation among the responses.
In a user interface evaluation study, the investigators typically design a few highly correlated questions for one evaluation criterion to reduce the effects of individual differences and randomness in participants' responses, e.g., NASA TLX (\citealt{hart1988development}), SWAT (\citealt{reid1988subjective}), etc.
The proposed MRMR model benefits from the quantification of the designed correlations among responses.
Therefore, it can serve as an analytical tool for HCI researchers and practitioners to investigate the individual differences.

On the other hand, the proposed MRMR method also provides good model interpretation because of the variable selection.
Based on the 95\% confidence intervals constructed from 50 estimates obtained from 50 random splits of data, we found that the selected variables for $Y_4$ include ``the average length of frequent eye movement trajectories", ``the alpha and gamma band of EEG signals", ``variance of mouse moving over/out duration", etc.
The variable ``average length of frequent eye movement trajectories" can reflect participants' different scanning paths.
Specifically, a higher value indicates continuous and fluent visual searching, and a lower value indicates that the participants may be confused with the visualization layout.
The selection of EEG-related variables aligns well with some research findings in literature.
For example, the variables ``alpha and gamma band of EEG signals" was reported to be significantly correlated with sustained attention during performing the visual searching tasks (\citealt{huang2007multi}).
Therefore, the estimated coefficients of these selected variables can help to understand how the visualization designs help participants concentrate on the contents.
Besides, the variable ``variance of mouse moving over/out durations" reflects the participants' visual search strategy of using the mouse to direct their attention, which is intuitively correlated with their total numbers of explored AOIs in different visualization designs.
These insightful understandings generated by the variable selection results of the proposed MRMR model can further improve the visualization designs.

\section{Discussion}\label{sec:summary}
In this paper, we study the online evaluation problem for interactive information visualization designs based on non-invasive measurements (i.e., wearable sensor signals).
This research aims at evaluating designs online catering to users' preferences and characteristics with mixed subjective responses by using non-interference measurements.
In this case, a joint model is able to improve the estimation and prediction results than modeling the responses separately when responses are correlated.
Therefore, a generalized linear model with mixed (i.e., continuous, counting, and binary) responses is proposed to quantify the correlation between mixed responses by using a graphical model.
The penalty terms are imposed on the negative likelihood function to encourage the sparsity in the estimated coefficient and precision matrices, such that the model is interpretable.
Both the numerical study and a user study in visualization evaluation show the advantages of the proposed MRMR method over the separate models.

The proposed MRMR model can be readily applied to the design and evaluation of information visualization in smart manufacturing, which can help novices understand complex process data sets and help experts generating insights from visualized data streams.
For example, the MRMR method can be directly applied to efficiently evaluate and improve the AR-based visualization platform proposed in \cite{chen2016iserc}.
Specifically, when manufacturing users (e.g., operators, engineers, and managers) are performing tasks by exploring data in AR visualization platform, their cognitive status can be quantified by MRMR model and thus supporting online adaptive visualization. Furthermore, the applications of MRMR model are not limited to evaluation of visualization designs. It can also be easily extended to support evaluation studies in human factors and ergonomics and human-computer interaction when multiple subjective responses are collected for in-depth understanding of human behaviors.
In addition, by taking advantage of the model structure of generalized linear models, the proposed method is able to accommodate other common types of response variables with exponential-family distributions.
For example, the negative inverse link function can be used if the responses follow exponential and Gamma distributions.
The logit link function is used if the responses are from Binomial distribution.
The derivation of the likelihood function and the objective in Section \ref{sec:model} can be obtained in a similar fashion.
However, the proposed modeling framework can not cover all distributions from the exponential family, such as the Chi-square distribution and Wishart distribution.
For such distributions, it is not straightforward to construct the link functions used in the generalized linear models.

There are a few directions for future work.
First, to better understand and optimize the human decision making process in interacting with visualization designs,
an online learning method for multi-step and multi-scale decision making will be studied.
Second, when some response variables inherently have very high dimensions compared to other responses, one can seek for the conditional models which are constructed by fitting high-dimensional responses conditional on other responses.
Such a conditional modeling approach would properly reflect the importance of the high-dimensional responses.
The techniques of modified Cholesky decomposition (\citealt{Kang2019On, Kang2020An}) can be potentially useful for dealing with the conditional modeling for high-dimensional responses.
Third, one can extend the proposed MRMR model to accommodate the missing data, of which the parameter estimation can still be addressed under the EM framework.
In particular, when different response variables have different sampling intervals, the sampling bias would potentially cause inefficiency and inaccuracy in parameter estimation.
To address such an issue, one solution is to treat the unsampled data points as the missing values. Then the proposed method can be extended under the framework of missing-data imputation where the EM algorithm is commonly used for dealing with missing data.
Another possible solution is to borrow strength from the seemingly unrelated regressions (SUR) (\citealt{zellner1962efficient}) to allow different sampling intervals for different responses.
However, the conventional SUR can only deal with multiple continuous responses.
It will be an interesting future research to integrate the proposed modeling technique with seemingly unrelated regressions for the mixed responses data.

\bibliographystyle{apalike}
\bibliography{reference}
\end{document}